# Radiative Transfer and its Impact on Determination of Thermal Diffusivity in Thin Films

O. Yu. Troitsky[†]
Institute for Non-Destructive Testing,
National Research Tomsk Polytechnic University, Tomsk, Russia

H. Reiss
Department of Physics, University of Wuerzburg, Wuerzburg, Gemany
harald.reiss@physik.uni-wuerzburg.de

**Abstract**

This paper extends the recently introduced Front Face Flash Method for extraction of thermal diffusivity of thin films to samples of small optical thickness. The paper discusses the principal question whether diffusivity is uniquely defined in case a heated ceramic, thin film sample is non- or only partly transparent to radiation. The paper applies radiative Monte Carlo and Two-Flux simulations to numerically solve the Equation of Radiative Transfer. Both methods are integrated in a Finite Element scheme, with conduction coupled to radiation, to investigate heat transfer under strongly transient and non-linear conditions. In this paper, the Front Face Flash method is successfully applied to $ZrO_2$ and SiC samples. It thus provides a promising tool to also investigate the thermal transport properties of thin film superconductors, but applicability still has to be demonstrated. The approach is particularly suitable for analysis of coupled conduction/radiation processes.

## 1 Survey

Like other ceramics, Y-stabilized $ZrO_2$ (YSZ) due to its high temperature and chemical stability, and by its superior hardness, is a candidate for highly wear-resistant surface coatings for protection of metallic or glassy substrates from corrosion or erosion, or as buffer layers in semi-conductor manufacture, or even as solid electrolytes in fuel-cells. Intensive heat and mass transfer is involved during preparation of YSZ-coatings like in plasma spraying on gas turbine blades, and similar intensive heat and mass transfer processes arise in chemical vapour deposition of amorphous carbon on combustion engine components. Also in their use phase, protective thin films like $ZrO_2$ may be subject to very high temperatures and temperature gradients. Accordingly, the thermal diffusivity of $ZrO_2$-, SiC-, $TiO_2$- and other coatings has to be measured, during their preparation and during ageing of the films. For these conditions, thermal wave technique and flash-methods are available for remote measurement of the thermal diffusivity.

Standard flash methods apply very short energy pulses, with length in the order of few nanoseconds. The methods work successfully under the following conditions:

(a) stationary "background" sample temperature,
(b) non-transparency (opacity) of sample or substrate material, which means signals like temperature variations, the "answer" of the sample to a short-time disturbance arrive at a detector solely by thermal conduction,
(c) laboratory conditions, like in the thermo-reflectance microscope.

Traditionally, isotropic transport properties, too, have been considered a prerequisite for successful application of thermal wave or flash methods. But isotropic conduction properties can rigorously be fulfilled only with some single crystals or in non-layered, thin film materials provided there is no significant phonon back-scattering from sample surfaces, or with pure metals, in other words, in some rare cases only. While isotropic thermal conductivity is typical for cubic crystals, we hardly ever achieve single crystals in industrially manufactured thin films. Approaches have been reported recently [1-3] that allow determination of diffusivity from flash experiments also in case of anisotropic thermal transport properties, or in layered materials, or for detection of disturbances like cracks or pores.

Condition (a) of the above neither may be fulfilled during sample manufacture nor during use phases. Plasma spraying of a protective coating on a gas turbine blade, for example, is performed by running quickly a beam (droplets of molten metallic or ceramic powder) delivered by a spray gun, line by line, over the substrate. This not only induces strong variations of sample temperature but coating thickness would increase rapidly during in-situ measurement, and crack formation may arise immediately behind the solidification front due to thermal contraction or thermo-mechanical mismatch between coating and substrate. Contrary to rapid film growth, the time interval, $\Delta t$, to reach a stationary (thermal) state, can extend to the order of seconds (it scales with $L^2/D$, with L the momentary thickness of the coating and D its thermal diffusivity). Thus stationary thermal conditions, a prerequisite for application of thermal wave technique, hardly can be achieved. The same applies to item (c) of the above if the sample is integrated in a running industrial process.

The thin film may either be partly transparent or non-transparent (opaque) to radiation. Clearly, the first case is not very suitable for application of thermal wave or of flash technique: The aim of a usual experimental setup is to determine the *solid thermal diffusivity*, and as a consequence, the applied method must respond *solely* to solid thermal conduction. Accordingly, it is item (b), the extinction coefficient of the material that must be very large, if film thickness is small.

Naturally, tempting assumptions like "opaque sample" have often been made in the literature, to simplify modeling heat transfer and interpretation of results. While diffusivity of metals and alloys or non-transparent dielectrics can be measured in a rather straightforward manner, materials and coatings of small optical thickness have mostly been investigated only when an irradiated extra covering layer was deposited that must be ultra-thin and opaque (the investigated material simply was "blackened" to incident radiation). But it is the covering layer that then acts as the secondary radiation source with its emitted, frequently black body, spectrum quite different from the spectrum of an incoming radiation beam.

The target under industrial conditions also might not be easily accessible, and there may be strongly scattering atmospheres like in a plasma spraying chamber where a lot of dust (particles that bounce back from the substrate) disturbs extraction of diffusivity from measurement of surface temperature (this is the reason why the present simulations start with definition of surface heat sources instead of modelling interactions between an incident beam and the sample front surface, see later for details).

The optical thin case accordingly remains to be solved for determination of thermal diffusivity, under coupled conduction and radiation conditions. For this purpose, the Front Face Flash method, recently introduced by the present authors, in particular applies measurement of surface temperature *distributions* instead of just the temperature of a single spot. Before the procedure is described in detail, we will first discuss the traditional Parker and Jenkins approach.

## 2 The Parker and Jenkins approach

In their frequently referenced paper, Parker and Jenkins [4], for a flat thin film sample under a transient surface heat source, successfully derived a solution of the thermal conduction problem: A single heat pulse is delivered from an arbitrary energy source to this layer. The point is that the heat pulse shall create solely a surface heat source and solely 1D thermal solid conduction, at least within the target area. Traditionally, this second condition is approached by keeping target very small to sample radius and with a large ratio of target area to sample thickness. But the penetration depth goes to zero only if the absorption coefficient of the thin film material is very large. If there are radiation transmission "windows" in some regions of the wavelength spectrum of the sample material, neither will penetration depth be infinitely small nor will energy transfer confined to solely solid conduction.

In case a perfect, solely 1D thermal conduction really would be achieved, and if also heat exchange with ambient (convection, radiation) can be neglected, the dimensionless rear surface temperature, $\Theta(L,t)$, on a flat sample of thickness, $L$, and at *any* lateral position, $x$, is obtained:

$$\Theta(L,t) = T(L,t)/T_{max}(L) = 1 + 2\sum(-1)^n \exp(-n^2 \pi^2 D t/L^2) \qquad (1)$$

Eq. (1) results from a series expansion of the solution to Fourier's differential equation; for details of its derivation see [4]. In a laboratory experiment, the $\Theta(L,t)$ is easily measured and compared with Eq. (1) to extract the diffusivity, $D$.

This ideal case neglects *any radiative transfer* in the sample (any heat transfer mechanism other than solid conduction). A well known exception of this case is diffusion-like radiative transfer. It is observed if the radiation extinction coefficient is very large and the mean free path of photons in the sample accordingly very small (this is the diffusion model of radiative transfer). However, in reality, sample thickness may be very small so that optical thickness is significantly limited. This is exactly the problem that shall be investigated in this paper.

For solely 1D solid conduction heat transfer, Figure 1 shows numerically calculated front and rear side temperature distributions; these results are obtained from a Finite Element procedure to solve Fourier's differential equation. Target radius equals sample radius in this simulation. Thermal equilibrium is achieved at about 1 s after start of heating.

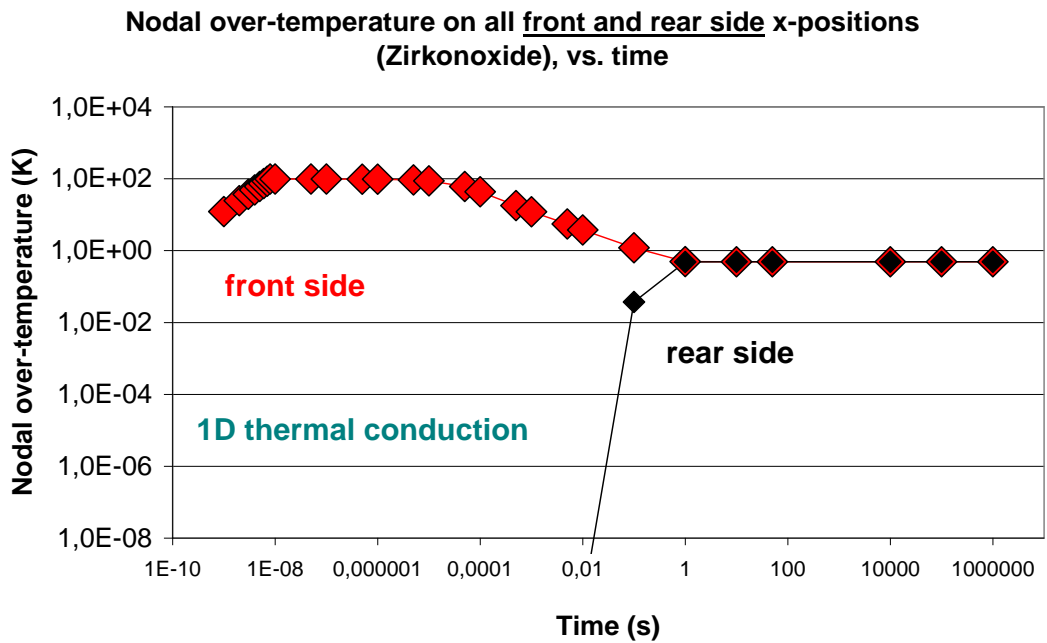

Figure 1 Nodal over-temperature of $ZrO_2$-sample, calculated vs. time, *under solely 1D solid thermal conduction* and adiabatic conditions. Co-ordinates, dimensions and symmetry conditions mentioned in the following are specified in Figure 3b: Sample (pellet) and target radii are $r_p$ = 120 mm, $r_t$ = 120 mm (target equals pellet cross section in this figure; results thus apply to *any* front or rear surface position). Sample thickness L = 1 mm, total length of heating period t = 8 ns, total deposited energy $Q_0$ = 64 Ws with heat flow density onto target area constant in time during the heating period (rectangular pulse); the calculations assume isotropic thermal conductivity and specific heat (both constant, independent of temperature). Results are given at central front (x = 0, y = 0) and rear surface positions (x = 0, y = 1 mm). Thermal equilibrium is observed at about 1 s after start of the irradiation (sample stagnation over-temperature under $Q_0$ is 0.488 K).

But if the target radius is very small, for example only 1/100 of sample radius (Figure 2a), thermal equilibrium is achieved much later, after several hours, and first deviations from a constant rear surface temperature will be seen (Figure 2b).

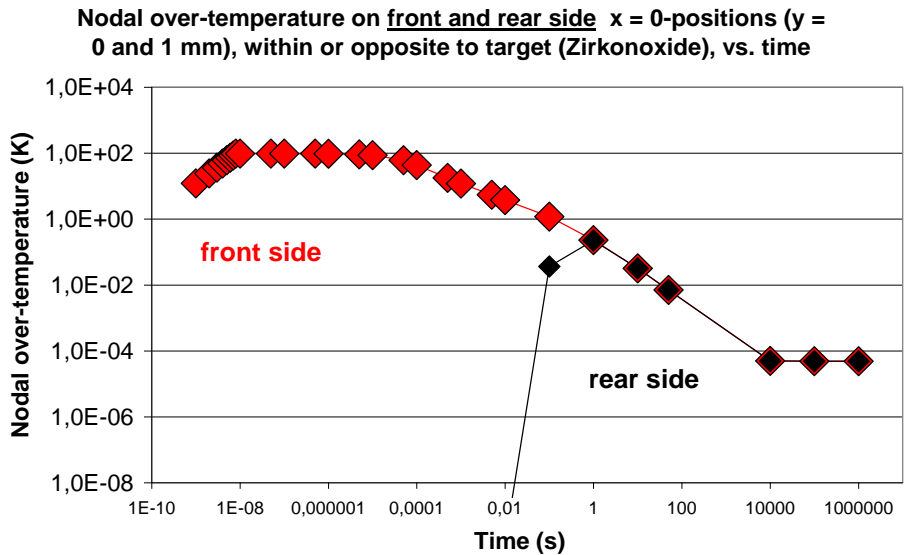

**Nodal over-temperature on <u>front and rear side</u> x = 0-positions (y = 0 and 1 mm), within or opposite to target (Zirkonoxide), vs. time**

<u>Figure 2a</u> Nodal over-temperature of $ZrO_2$-sample, calculated as before (Figure 1, solely solid conduction) but with target very small against sample (pellet) radius, $r_p$ = 120 mm, $r_t$ = 1.2 mm. As before. all co-ordinates and dimensions are specified in Figure 3b. Total length of heating period again t = 8 ns, total deposited energy $Q_0$ = 0.0064 Ws (rectangular pulse, as before). Results are given at central front (x = 0, y = 0) and rear surface positions (x = 0, y = 1mm). Thermal equilibrium is achieved not before several hours after start of irradiation (sample stagnation over-temperature is 0.488 $10^{-4}$ K).

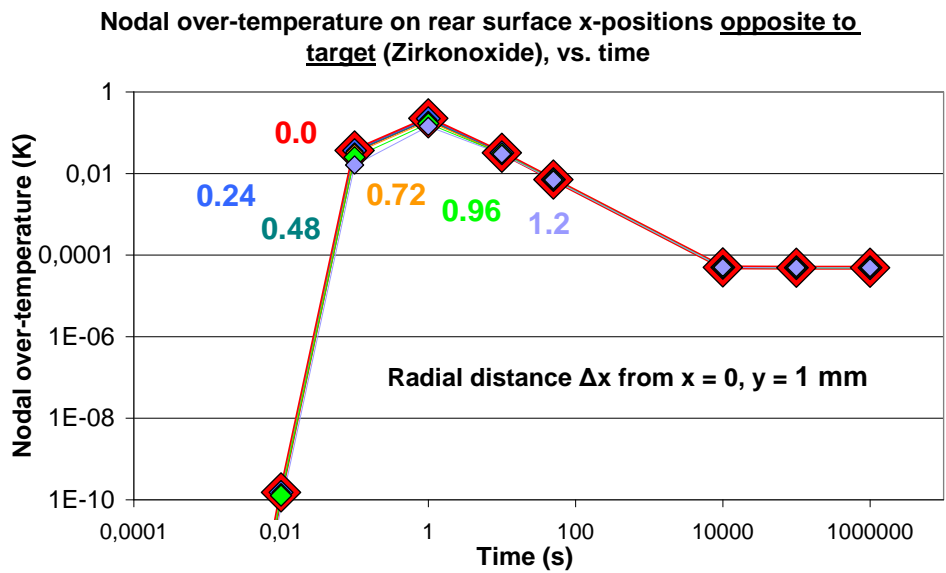

**Nodal over-temperature on rear surface x-positions <u>opposite to</u> target (Zirkonoxide), vs. time**

Radial distance Δx from x = 0, y = 1 mm

<u>Figure 2b</u> Nodal over-temperature of $ZrO_2$-sample, calculated as before (Figure 2a), with target very small against sample (pellet) radius. Results are given at different radial distances, Δx (mm), from central *rear* surface positions (x = 0, y = 1mm).

While the original Parker and Jenkins model has been modified in the literature to account for deviations from 1D conduction or for heat exchange with ambient, the essential problem remains, i. e. when radiation would be coupled to conduction. This is inevitably the case when extinction coefficient, E, or sample thickness, L, are small so that the optical thickness τ = E L would be limited to τ < 15; for an explanation of this numerical limit see [5] or recent work by the authors.

The numerical simulations, to investigate the coupled conduction/radiation problem in thin films, that we have performed include

(a) calculation of transient front and rear surface (over-) temperature and their excursion with time; these will be obtained from Monte Carlo- and Two Flux model radiative calculations, which are combined with a Finite Element conduction heat transfer model,

(b) application of the Front Face Flash method recently introduced by the authors to extract thermal diffusivity; the method relies on transient surface temperature *distribution*s instead of considering only a single temperature such as Θ(L,t) in Eq. (1).

(c) The present paper, in particular following item (b), presents a check of the accuracy of the Front Face Flash Method (see below, Sect. 5): In a first step, it applies transient temperature distributions obtained from Finite Element simulation that are performed with experimentally known $ZrO_2$ and SiC thermal diffusivity, solid conduction and extinction coefficients (but in principle could be of arbitrary values). The resulting temperature distributions in the second step are used to extract the diffusivity, D, using Eq. (3), an expression derived by the late co-author (Oleg Yu. Troitsky) from solutions of the Fourier differential equation.

### 3   Modeling of radiative heat transfer coupled to solid conduction

The Monte Carlo simulation described below shows that creation of a heat source, not located below but just at exactly the sample surface (the Parker and Jenkins assumption), cannot be realized with existing (standard) materials; a perfect thermal mirror would have to be constructed to realize the condition "zero thickness an absorptive surface layer" for the derivation of Eq. (1); this is not possible.

The assumption "opaque" sample in addition has led many authors to believe in an opaque solid there are no radiation contributions to heat flow at all, and that corresponding source terms in the Equation of Radiative Transfer (ERT), compare e. g. [6], Sects. 14, 15, 19 and 20, can be neglected. However, opacity only means that an *original* ray is extinguished over very short distances. This does not mean there is *no* radiation in the material. The only concession that can be made is to admit that radiation in the interior of opaque materials, at an arbitrary position, (x,y), is of *local* origin only, i. e. comes just from a tiny volume of radius $l_m$ (the mean free path of photons) around the position (x,y), but from no-where else; we then speak of radiation as a diffusion process.

Radiation diffusion models have been reported in the literature for this situation (see e. g. [6], Sect. 15-4, or for a short description [7], Sect. 3.4). Such models accordingly treat propagation of radiation as a conduction process, which enormously simplifies calculation of coupled conductive/radiative heat transfer.

The majority of reports available in the literature also neglects that flash experiments lead to radiative *non-equilibrium*, which means radiation not only would be absorbed but also might be scattered and perhaps escape from the sample; this definitely has impacts on its overall energy balance, on sample temperature and thus on measured diffusivity.

If the optical thickness is large (with the above mentioned τ = 15 as the lower limit), a Monte Carlo approach can be applied, while for small optical thickness, a radiative Two Flux-model is given preference, see below. Both models serve for calculation of absorbed and (anisotropically) scattered radiation. Components of the ERT are illustrated schematically in Figure 3a. While its principal solutions exhaustively have been discussed in the literature, see e. g. [6-9] for in-depth descriptions, the solutions apply mostly to ideal cases (extinction properties independent of wavelength and temperature, isotropic scattering, constant and isotropic solid conductivity and specific heat). None of these is fulfilled in reality.

In particular, isotropic scattering is rather an exceptional case. However, the solutions at least allow to demonstrate *tendencies* how coupled solid conduction/radiation heat transfer would have impacts on temperature distribution in a thin film sample, and how the corresponding variations would alter the extracted thermal diffusivity. Simultaneously to solutions of the ERT, the equation of conservation of energy, both applied to the present problem, has to be solved; in the present paper, this is done by the Finite Element simulation.

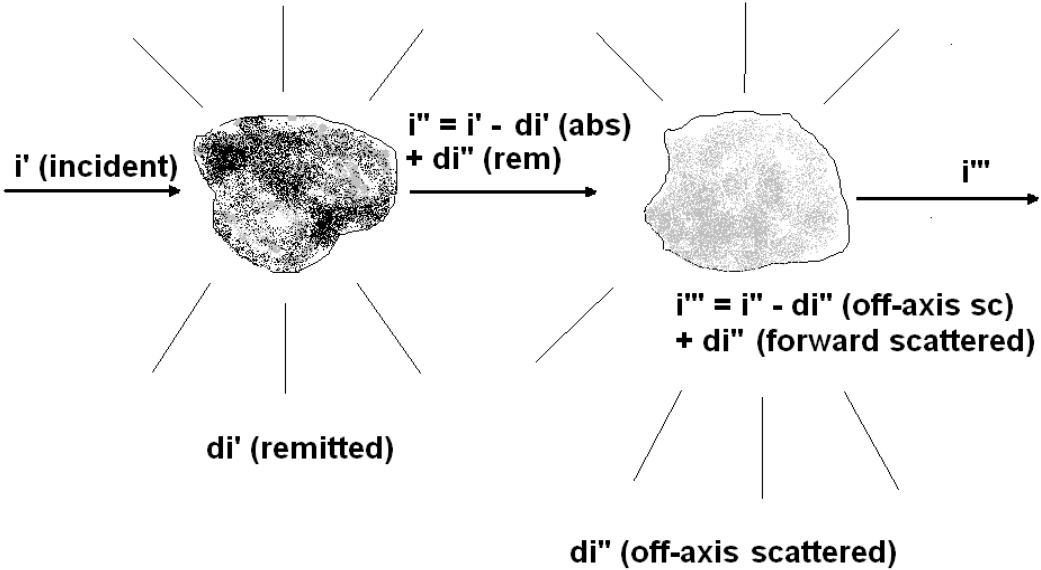

Figure 3a Components of the Equation of Radiative Transfer (ERT; schematic); solutions of the ERT in the present paper are approximated by a Monte Carlo-method and a radiative Two Flux-model. In both models, all parameters describing propagation of a bundle and creation of volume sources are treated as random variables. The figure is taken from [10].

All details of the Monte Carlo approach applied to determine the radiative contributions to the total, transient heat transport problem in thin films, have been reported in [10] and will not be repeated here but a short description is given in the following.

**Step 1a: Monte Carlo method**
Figure 3b shows a longitudinal horizontal section of a cylindrical thin film pellet. Discrete co-ordinates (i,j) are used to identify plane "area elements" (i ≤ 50, j ≤ 20) for the Monte Carlo simulations in the pellet. The area elements later will each be filled with a large number of proper "finite (FE) elements"; the corresponding FE-mesh is not shown in this figure. Continuous co-ordinates (x,y) serve to solve Fourier's differential equation in the Finite Element procedure. Volume elements (cylindrical shells), with same (i,j) co-ordinates, are generated by rotation of the meshed area elements against the symmetry axis (x = 0), and so are the finite elements..

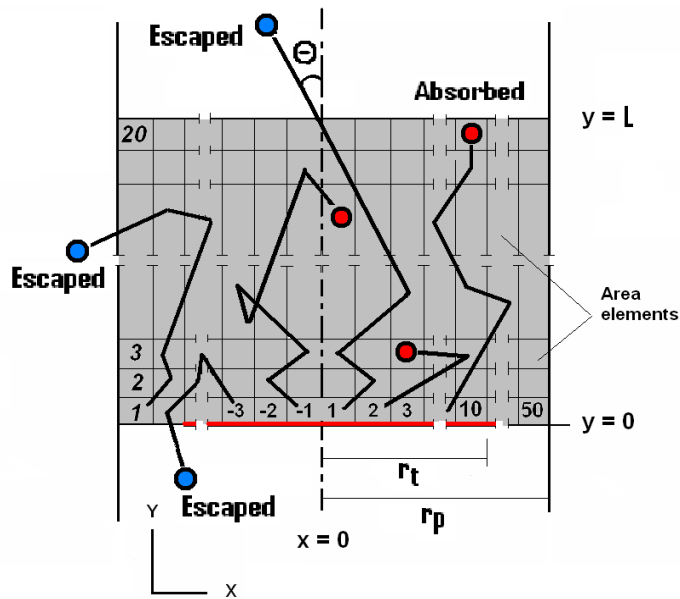

Figure 3b Horizontal section of a cylindrical pellet; the section includes the symmetry axis (x = 0, thick dashed-dotted line). Right and left halves of the pellet are divided each into a number of 1000 area elements (i ≤ 50, j ≤ 20) that are meshed with in total about $8 \cdot 10^4$ plane finite (FE) elements (not shown in this figure). Area elements (i,j) are indicated dark-grey, and radiation bundles by thick black lines. Rotation of the area elements against the axis of symmetry (x = 0) generates volumes elements (cylindrical concentric shells). Pellet radius, $r_p$, and thickness L, are 120 and 1 mm, respectively. Target radius is $r_t$ = 120 mm in Figure 1, 1.2 mm in Figure 2a,b, 15 mm in Figures 5a,b and 8a,b and 9a,b, and again 1.2 mm in Figs. 6a,b and 7a,b. The target, identified by the horizontal thick red line, is irradiated or otherwise flash-heated during 8 ns (rectangular pulse), with uniform total energy depositions, $Q_0$ (Ws), onto the target elements (volume elements, cylindrical shells occupying the first row, j = 1, of the discrete co-ordinates i,j); since their cross sections increases with distance from x = 0, thermal power (W/m$^2$) decreases with larger co-ordinates, x. Scattering angle is denoted by θ. Bundles either are absorbed/remitted (red circles) or scattered and may escape from the sample (blue circles, index "Escaped") after a series of absorption/remission or scattering interactions with the sample material. The scheme is used in both Monte Carlo and Finite Element (FE) calculations.

The front faces (y = 0) of part of the cylindrical pellet elements serve to define the target plane, (i ≤ 10, j = 1). Each of the target elements is heated by a thermal load, $Q_0$ (Ws). The physical origin of this initial source (the transient disturbance of an initially homogeneous temperature distribution in the pellet) in principle is arbitrary. Examples are absorbed radiation from a single short laser pulse, or an electron beam (like in the paper by Parker and Jenkins), or even a surface heater.

Target or shell elements (and, after meshing, their corresponding finite elements) accordingly serve for definition of surface heat sources that initialise the subsequent thermal, solid conduction plus radiation transport processes. In this way, we need not model an initial radiative or electron beam interaction with the target.

In cylindrical geometry, radius of the elements (distance from the x = 0 axis) increases, and so does their individual cross section, A. All target elements shall receive the same absolute amount, $Q_0$ (Ws), of thermal energy, constant within t ≤ 8 ns. The thermal power density, $(dQ_0/dt)/A$ (W/m$^2$) onto the target surface elements thus decreases with increasing co-ordinate, i. At the central position, it is maximum. The load profile accordingly is not Lambertian but depends on the square of the corresponding i- (or x-) co-ordinates.

The disturbance not only emits a thermal wave (conductive heat flow emerging from positions y = 0 within the target area) are but in the Monte Carlo simulation also is the source of a large number of radiation beams (bundles), with randomly distributed foot-points (i ≤ 10, j = 1) and the subsequent absorption/remission or scattering events at random interior positions (i ≤ 50, j ≤ 20) of the sample. The foot-points oscillate over the target area. Once emitted, the beams, parallel to conduction, by absorption create randomly located, internal volume heat sources, $Q_V(i,j)$, at all simulated times, t. More beams (in the language of Monte Carlo simulations: more bundles) are remitted from these volume elements (i,j) to create new sources, with their distribution and magnitudes again simulated in the Monte Carlo model.

Mean free path, $l_m$, of radiation emitted from the target area and from all internal positions (i ≤ 50, j ≤ 20), scattering angles, $\Theta$, and albedo, $\Omega$, of single scattering determine the positions of the radiative volume sources, $Q_V(i,j)$. After each absorption/remission event along the zig-zag path of a bundle, magnitude of $Q_V(i,j)$ decreases until the bundle energy is completely exhausted, or when it escapes from the sample.

Radiation bundles escaping from the sample are schematically indicated in Figure 3b (index "Escaped"). They are taken into account in the Monte Carlo method, to fulfil conservation of (radiative) energy. The number of escaping bundles is not negligible but the analysis is confined to solely the *internal* heat transfer processes and their competition among each other.

The scattering phase function is explained in the Appendix (see the corresponding spider diagram; it indicates strong forward scattering).

The total number of bundles in the Monte Carlo simulations is at least $N = 5 \cdot 10^4$; this number proved to be sufficiently large, compare Figures 3 and 4b in [10] and Figure 4 below.

Within real materials, occurrence of absorption/remission and, in particular, scattering interactions is inevitable (a variety of crystal and lattice defects, grain boundaries, cracks and pores) are responsible for scattering. Thus heat transfer in any real sample, under a *localised* heat source like in flash experiments*, cannot be one-dimensional* if the material is not perfectly non-transparent (optical thickness $\tau \to \infty$), in *all* spatial directions. If the optical thickness is very large, a thermal detector that, for example, receives transmitted radiation from the corresponding rear sample surface neither responds to an energy flow from just one direction through the sample nor can it distinguish initial diffuse from initial beam radiation sources directed onto the sample front surface; even, in an absorbing material, it cannot distinguish the wavelength of incident radiation from that of the original source (this would be possible only if the material is solely scattering and under elastic collisions.

For determination of temperature excursion with time in a conductive solid, Carslaw and Jaeger [11] showed that an initial temperature distribution is equivalent to a distribution of instantaneous, initial heat sources. Conversely, once radiative volume power sources, $Q_V(i,j)$, have been determined in the Monte Carlo simulation, this distribution is equivalent to an initial temperature distribution within the pellet. Provided the optical thickness is large, it then makes sense to treat the whole coupled conduction/radiation problem as a *conduction* process, using an appropriately defined radiative (diffusion) conductivity.

**Step 1b: Radiative Two Flux model**

Application of the Monte Carlo simulation described in Step 1a to also samples of small optical thickness, τ = 1, becomes extremely laborious: After initial interactions between bundle and constituents of the pellet (solid particles, cracks, pores) at positions (i,j), more interactions of the same (residual) bundle emitted from these positions either occur with other solid particles or with other obstacles at positions (i', j'). Or the bundle would immediately escape to the ambient. The (i',j') no longer are located very close to the (i,j); this would be the case only if the optical thickness is large. The optical thickness by its definition, τ = L/$l_m$, with $l_m$ = 1/E, on a statistical basis, the mean free paths between two interactions, is identical with the number N of interactions in the pellet. In the present case, N = 1 if τ = 1; there is, accordingly, and still on the statistical basis, in the extreme case *no* secondary absorption/remission or scattering event at positions (i', j') within the sample.

Each volume element, under the condition τ = 1 and *if* it is hit by a beam becomes an absorbing/remitting or scattering obstacle (or, if it is not hit, a thermal emitter simply because it is heated by conductive heat flow). In this way, radiative *exchange* over extended distances (not a diffusive, radiative conduction process), is initialised, with all distant neighbours.

In contrast, in the optical thick limit (τ →∞), the radiative interaction, as a diffusive *transport* mechanism, again includes all elements but extends over only very small distances (as mentioned, explicit modeling of each of the huge number of these diffusive interactions can be circumvented by introduction of radiative conductivity).

For an adequate approximation of the optically *thin* limit, time integration steps therefore would have to be chosen very small in order to re-calculate temperature of all volume elements before a new distribution of power sources and their emissions can be determined. The overall procedure then requests a close series of Monte

Carlo calculations and intermittently performed finite element calculations to determine emissive vs. received radiative power and temperature of each volume element. Although the procedure in principle is straight-forward, this sequence would enormously increase computation time, even if the axial symmetry of the pellet is exploited. Yet the Monte Carlo method, though extremely laborious, also under these conditions finally delivers reliable surface temperature distributions (see below, Figures 8b and 9b).

Because of the increased mean free path, it is to be expected the distribution of power sources (and thus of temperature) within the sample becomes more uniform (like the distribution of bundles leaving the rear surface in Fig. 4). Volume sources will no longer be observed only close to the sample front side, and since a considerable part of the bundles is lost (escaped). These expectations are confirmed in [10], Figs. 17 - 22; also see the discussion below, Sect. 5.2.

An alternative of practical value for the case τ = 1results from application of a radiative Two Flux model:

The Two Flux-model yields radiation transmission coefficients, $\zeta = q_+(x,y=L)/q_+(x,y=0)$ from ratios of positively oriented radiative flux, $q_+(x,y)$, with L the sample thickness; the corresponding. oppositely oriented flux $q_-(x,y)$ would serve for calculation of reflection coefficients from the ratio $q_-(x,y=0)/q_+(x,y=0)$, but this is of minor importance for the internal transport processes investigated here. During given time steps, energy $Q(x,y) = \zeta\, q_+(x,y=0)$ is delivered either to a volume element (i,j) located inside the pellet, or the flux leaves the pellet. Taking into account the albedo Ω of single scattering, like in the Monte Carlo simulations, the energy $Q_{V,abs}(i,j) = (1 - \Omega) Q(i,j)$ is absorbed while $Q_{V,sc}(i,j) = \Omega\, Q(i,j)$ is scattered at the position (i,j).

Derivation of the transmission coefficient, ζ, in the Two Flux model (a series of exponentials in E, Ω and L) is straight-forward and is described in standard volumes on radiative transfer (like [6, 8-9], all these by no means are old-fashioned; a short description can also be found in [7], Sect. 3.1). Higher order approximations than provided by the Two Flux Model, e. g. the method of discrete ordinates, compare [12, 13], are recommended, in case scattering is very strongly anisotropic.

In its most simple form, the Two Flux-model applies to a cold medium. Nodal over-temperature, under coupled conduction and radiation, has been limited in the present simulations to values below 20 K, see below (blue and green symbols in Figures 6a,b and 7a,b).

**Step 2: Finite Element model**

Literature values for thermal conductivity, specific heat and density will be taken as input (like proper "experimental" data) to the Finite Element calculations. The Front Face Flash method then will be applied to the thus generated temperature distributions and their excursion with time, to finally extract the diffusivity. In order to verify the method, the extracted diffusivity is compared with the original input data.

Radiation absorption, remission and scattering, in the interior of the solid, proceed by velocity c/n of light (n the refractive index); propagation of a thermal wave, by solid conduction/diffusion only, is much slower, by orders of magnitude. Absorption and remission of radiation emanating from the target surface (x ≥ 0, y = 0) and from the $Q_V(i,j)$ at interior positions accordingly can be considered as *initial* conditions (like an initial temperature distribution) to the subsequently treated thermal conduction problem (this reflects the above mentioned theorem of Carslaw and Jaeger [11]).

Thermalisation of the sources $Q_V(i,j)$ is calculated using a standard finite element (FE) program (Ansys, Release 16, with plane finite elements). Mapped meshing is applied in the FE calculations; we have up to $8 \cdot 10^4$ nodes and plane elements and the corresponding number of volume elements. For simplicity, thermal diffusivity is provisionally taken as constant, i. e. independent of temperature. The proper impact of radiation might be disguised if temperature dependent diffusivity was used, or if convection or radiation exchange with ambient would be integrated (though both conditions can be considered straightforward in the FE-calculations).

The complete data input (conduction and radiation) is given in the Appendix.

## 4      Surface temperature distributions

While Rosseland mean extinction coefficients were applied in [10], to account for spectral dependency of the extinction coefficient, E, the present analysis is restricted to constant E (independent of wavelength and temperature), again to keep description of the method as simple as possible (and tolerable).

Magnitude of E = A + S of poly-crystalline YSZ (with A the absorption coefficient taken from an experimental spectrum, and S, the scattering coefficient, from Mie-theory of scattering) was estimated in [10], Sect. 4.1 (and in [14] for a ceramic superconductor). A Rosseland mean extinction coefficient, $E_R$ = 1.696 $10^4$ [m$^{-1}$], results from experimental spectral YSZ extinction coefficients so that a range $10^4 \leq E \leq 5 \cdot 10^4$ 1/m applies approximately to the thermal spectrum of this and other poly-crystalline ceramic solids, in the present range of temperatures (300 ≤ T ≤ 400 K).

## 4.1    Samples of large optical thickness

Accuracy of the Monte Carlo-approach can be checked e. g. by calculated angular distribution of the directional intensity, I', of absorbed/remitted and scattered radiation; the intensity is in Figure 4 represented by the directionally emitted, residual number of bundles.

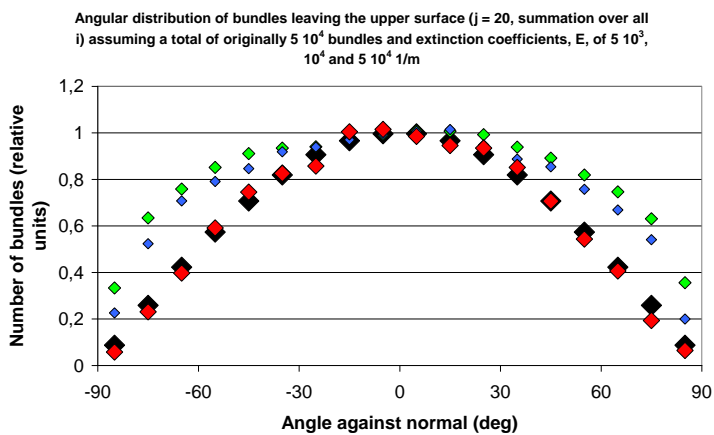

Figure 4 Angular distribution obtained from the Monte Carlo-simulation of the residual number of (originally emitted) N = 5 $10^4$ bundles leaving the outermost layer, j = 20 (y = 1 mm) of the cylindrical pellet (rear sample surface); data are plotted against the emission angle, Θ, against normal (compare Figure 3b). The distribution (obtained after summation over all horizontal co-ordinates (i, oriented parallel to the continuous co-ordinate, x) is calculated for $ZrO_2$ using extinction coefficients, E [m$^{-1}$] = 5 $10^3$, $10^4$ and 5 $10^4$ (solid light-green, blue and red diamonds, respectively; these correspond to optical thicknesses τ = 5, 10 and 50). Solid black diamonds indicate the theoretical, diffuse cos(Θ)-distribution given by the Lambert law. The larger the extinction coefficient (or the optical thickness), the better is the angular distribution of the bundles leaving the sample from its rear surface represented by a diffusely radiating surface.With inreasing extinction coefficient (or optical thickness), their distribution

shown in Figure 4 approaches more and more closely the well-known Lambert cosine law. Note that the distributions refer to emissions from the rear sample surface.

Figure 5a,b shows calculated front and rear side over-temperature obtained from the combined Monte Carlo- and Finite Element method (the results are taken from [10], with pellet and target radii of $r_p$ = 120, $r_t$ = 15 mm). As is to be expected, a uniform temperature distribution on either front or rear sample surface cannot be confirmed. Instead, a strong hot spot shown in Figure 5a,b appears in the target plane (front and rear surfaces) near the position (x = 0, y ≥ 0).

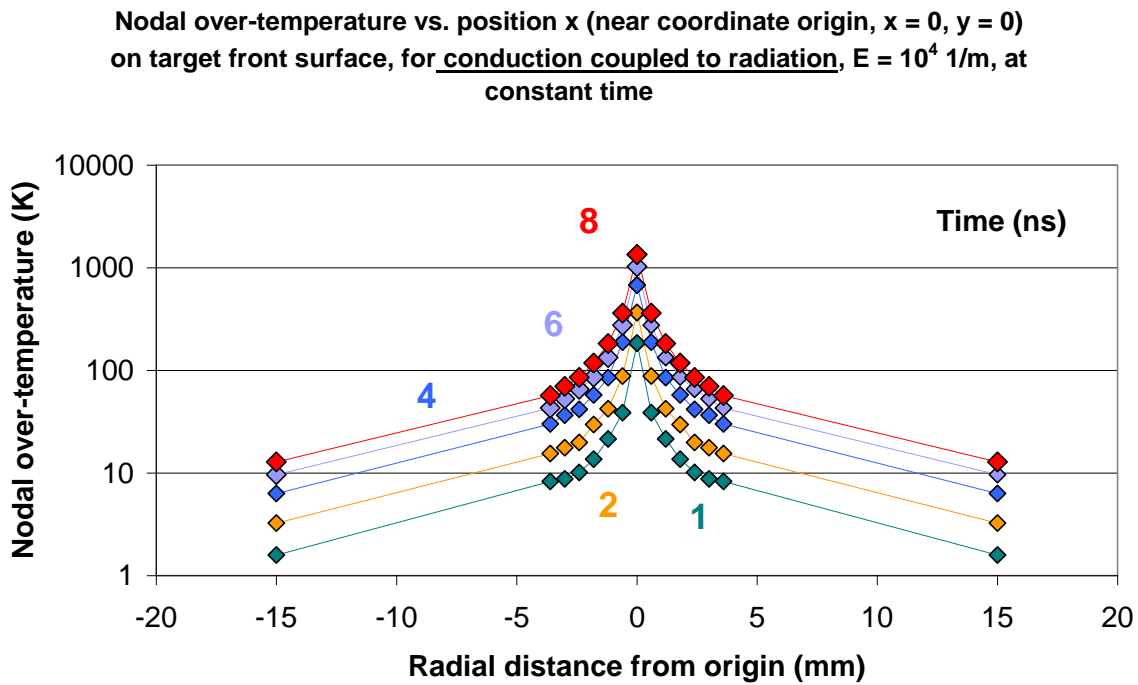

Figure 5a Nodal over-temperature of ZrO$_2$-sample on its *front* surface, calculated from the Monte Carlo-simulation that is integrated into the FE procedure, for *solid conduction coupled to radiation,* with target radius small against sample (pellet) radius ($r_p$ = 120, $r_t$ = 15 mm, this may reflect experimental, industrial laser flash or otherwise heated targets), under adiabatic conditions (no convection or radiative interaction with ambient) and for large optical thickness, τ = 10. Total length of heating period is 8 ns, total deposited energy $Q_0$ = 1 Ws. Results are given at different radial distances Δx (mm), within the target area, from central surface positions (x = 0, y = 0) and for times 1 ≤ t ≤ 8 ns. Albedo of single scattering Ω = 0.5; strong anisotropic forward scattering (particle size comparable to incoming wavelength), isotropic conductivity and specific heat (independent of temperature). The temporarily existing hot spot at x = 0 arises from distribution of incident thermal power (compare text), from strong forward scattering and from the large optical thickness ($τ = E\, L = 10$, with $E = 10^4$ 1/m, the extinction coefficient, assumed as constant, independent of wave-length); all these strongly concentrate absorption/heating events to forward (y) positions close to target. Data are from [10].

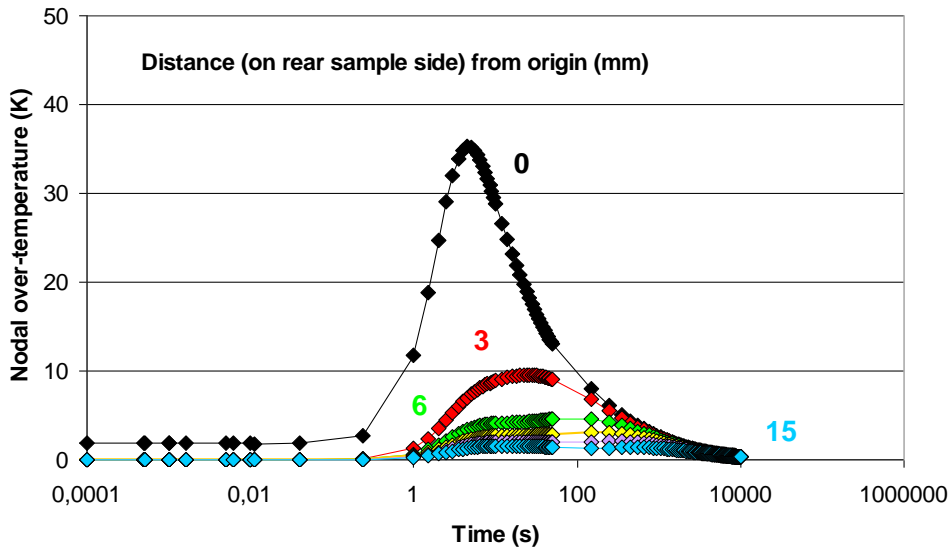

Figure 5b Nodal over-temperature of $ZrO_2$-sample vs. time; same calculation as in Figure 5a but results are given on the *rear* sample surface at different radial distances, $Δx$ (mm) from the origin ($x = 0$, $y = 1$ mm), i. e. within the area just opposite to target area ($r_t = 15$ mm) using the Monte Carlo simulation integrated into the FE procedure, for *solid conduction coupled to radiation*. The temporarily existing hot spot at $Δx = 0$ mainly results from the corresponding strong temperature peak on and near the front surface; it reflects the given distribution of thermal power but also strong anisotropic forward scattering and the large optical thickness; all of these concentrate absorption/heating to forward (y) directions. The figure is taken from [10].

The origin of the hot spot is three-fold: First, it results from the previously explained radial dependence of the volume element cross sections (circular, concentric shells). They all receive the same absolute amount, $Q_0$, of energy, but the shells are exposed to strongly different heating power, $(dQ_0/dt)/A$, per unit area, A. Since the cross section of element $i = 1$ (or the volume of this element) is minimum, surface temperature of this element (strictly speaking, temperature at $x = 0$, $y = 0$) becomes maximum (Figure 5a).

The impact of solely the distribution of thermal sources in the target plane on surface temperature can be seen below, solid red circles in Figure 6a: If solid conduction is the only transport mechanism, a strong hot spot arises at position (x = 0, y = 0) also without any radiation contribution. A hot spot is also observed at the rear sample surface (below, Figure 6b).

Second, if there is coupled conduction/radiation, the hot spot in Figure 5b (rear surface, positions opposite to target plane), with smaller magnitude, reflects the strongly anisotropic, forward scattering events experienced by all $5\cdot 10^4$ radiation bundles once they are emitted from (a) the target and (b) when they are absorbed/remitted and strongly forward scattered by all elements (i,j) hit by the bundles.

Third, beams scattered or emitted in directions off normal have to run over extended distances to volume elements (i',y') located anywhere in the sample; this means these beams will loose more energy and thus are quickly extinguished. Elements (i,j) located at directions off normal and at large co-ordinates, j, and because of the low temperature rise receive little radiation originating from the initial disturbance.

The traditional approach (data taken in principle at *any* rear sample surface position, Figure 1), under coupled conduction/radiation conditions thus is not applicable.

## 4.2 Small optical thickness

Figure 6a,b shows calculated front and rear surface over-temperatures at central pellet positions. Sample (pellet) and target radii are $r_p$ = 120 mm, $r_t$ = 1.2 mm, the same as in Figure 2a,b. Results are strongly different when considering solely solid conduction (solid circles) and conduction coupled to radiation (solid diamonds), now obtained from application of the Two Flux-model. Over-temperatures are given for strongly absorbing (albedo $\Omega$ = 0.01) and strongly (and anisotropically) scattering materials properties ($\Omega$ = 0.99).

While temperature excursion on the front surface is structurally similar in both cases ($\Omega$ = 0.01 or 0.99, Figure 6a), but with substantially differing magnitudes, it is also *structurally* strongly different on the *rear* sample surface (solid blue and solid dark-green diamonds, Figure 6b).

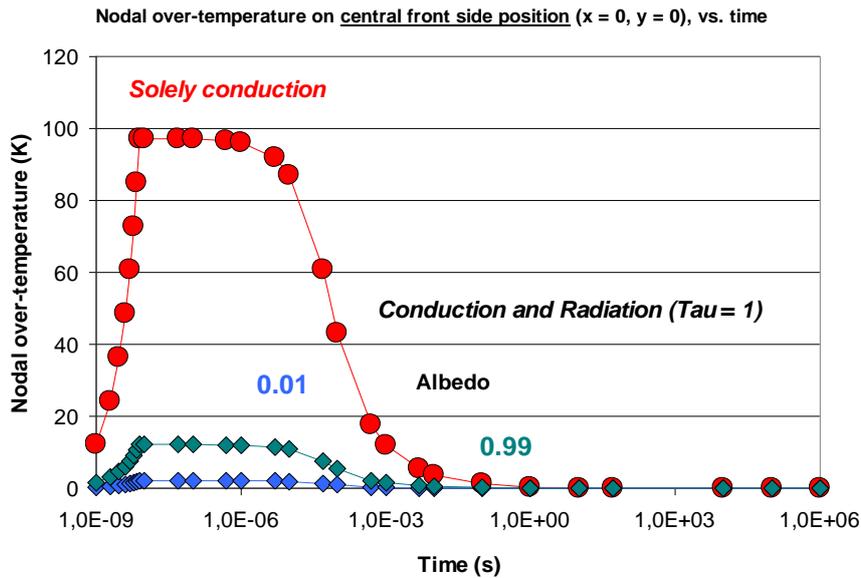

Figure 6a Nodal over-temperature of $ZrO_2$-sample on its *front* surface, calculated from the Two Flux-Model integrated into the FE procedure, for solely solid conduction (solid circles) and for solid conduction coupled to radiation (solid diamonds) with target very small against sample (pellet) area, under adiabatic conditions and for *small* optical thickness, $\tau = 1$. Pellet and target radii are $r_p = 120$, $r_t = 1.2$ mm. Results are given in dependence of time at central front position ($x = 0$, $y = 0$) and for albedo of single scattering $\Omega = 0.01$ (strong absorption) and 0.99 (strong scattering); scattering again is strongly anisotropic (forward directed). Length of total heating period and deposited energy $Q = 1$ Ws as before, isotropic conductivity and specific heat (independent of temperature). No significant hot spot arising at $x = 0$, except in case of pure conduction, which reflects solely the distribution of heat sources on the target area. Radiation and its absorption/remission and scattering quickly distributes energy deposited on the target area ($y = 0$) to interior sample positions ($y > 0$); this redistribution strongly reduces the calculated front side surface temperature, in relation to the solely conduction case.

In Figure 6b, the difference between solid circles and diamonds, in particular the peak in the series of solid red circles at about $t = 1$ s, illustrates the strongly differing propagation velocities of conduction and radiation in the sample material. Note also the coincidence of the peak values (solid red circles and solid green diamonds) at $t = 1$ s: Strong scattering reduces height of the dark-green symbols; this reflects escape losses that scale down absorption by elements on the rear surface and thus the surface temperature.

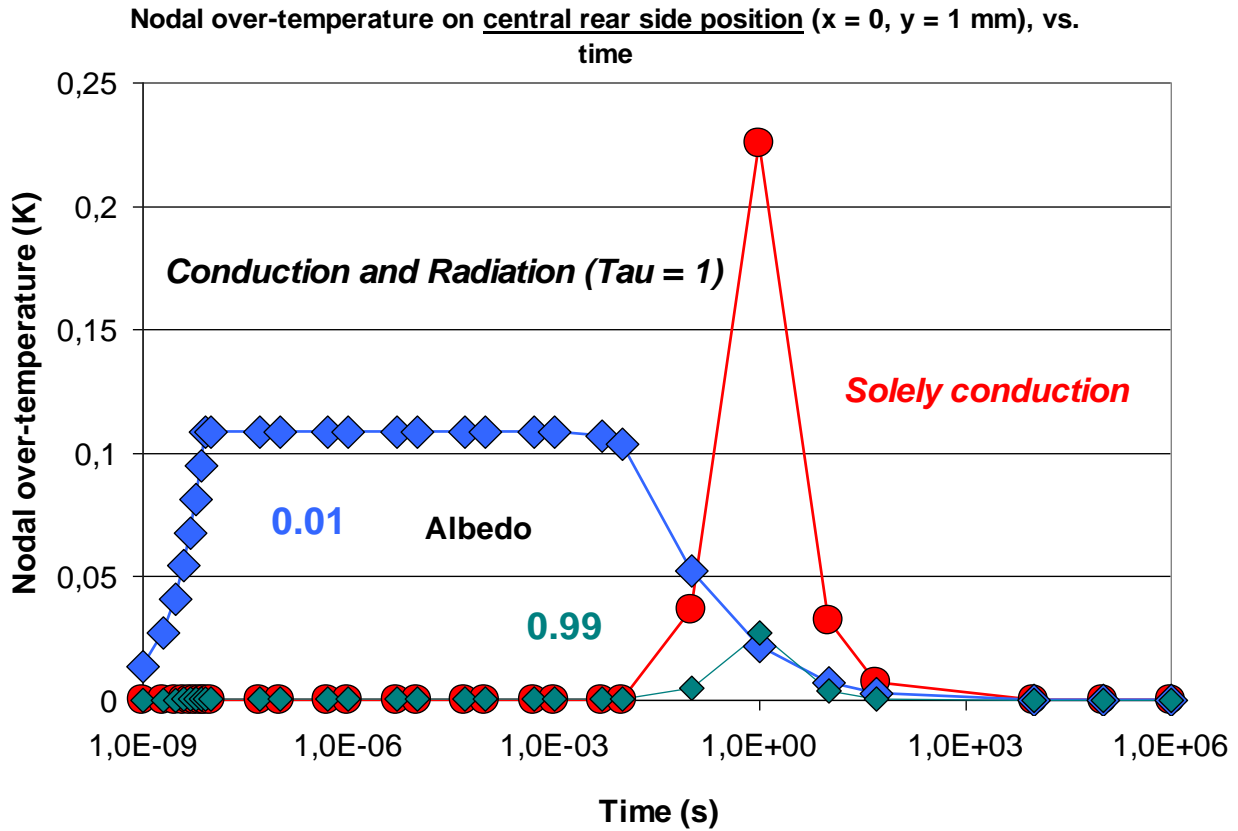

Figure 6b Nodal over-temperature of $ZrO_2$-sample on its *rear* surface, calculated as before (Figure 6a) from the Two Flux-Model integrated in the FE procedure, for solely solid conduction (solid circles) and for solid conduction coupled to radiation (solid diamonds), with target very small against sample (pellet), under adiabatic conditions and for *small* optical thickness, τ = 1. Results are given in dependence of time at central rear side position (x = 0, y = 1 mm). The strong peak (solid circles at about t = 1 s) illustrates the different velocities by which the disturbance (heating of the target area) proceeds in the sample material (propagation of radiation proceeds much faster than solid thermal diffusion). At times t < 1 ms, strong absorption (Ω = 0.01) increases sample temperature while large albedo (Ω = 0.99, strong scattering) is responsible for increasing radiative (escape) losses.

Figure 7a,b shows that like in Figure 5a,b surface temperature distributions are not at all homogeneous, which means the Parker and Jenkins approach or its upgrades, as far as they result in homogeneous surface temperature distributions, again are not applicable. But this is clear from the outset: No vanishingly thin absorptive surface layer can be created in a sample of small optical thickness, one of the essential assumptions of the Parker and Jenkins model (a thin absorptive layer would just shift the same problem to positions y > 0).

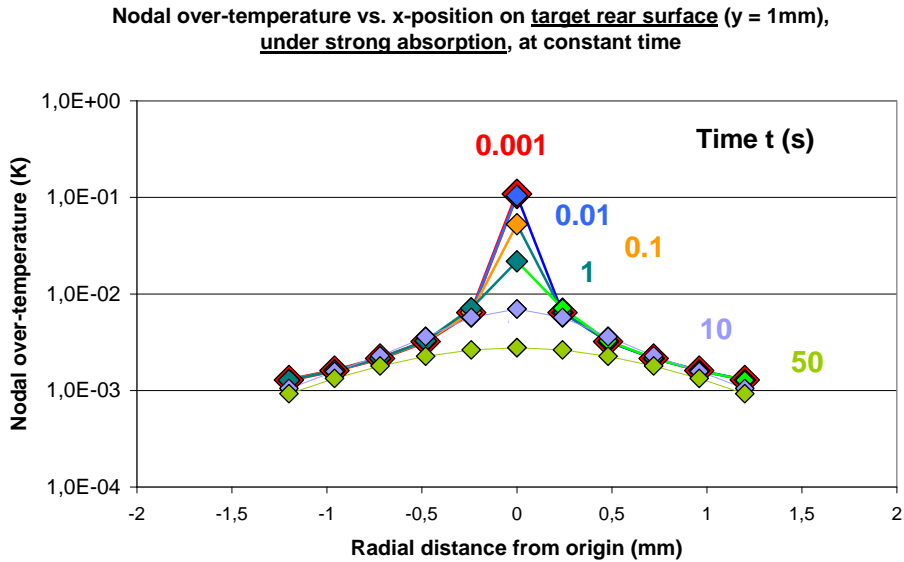

Figure 7a Nodal over-temperature of ZrO$_2$-sample on its *rear* surface, calculated as before (Figure 6a) for coupled conduction/radiation, here under *strong absorption*, using the Two Flux-Model integrated in the FE procedure, and for *small* optical thickness, τ = 1. Results are given at different radial distances, Δx (mm), from the origin (x = 0, y = 1 mm) within target and pellet area, at given times, t.

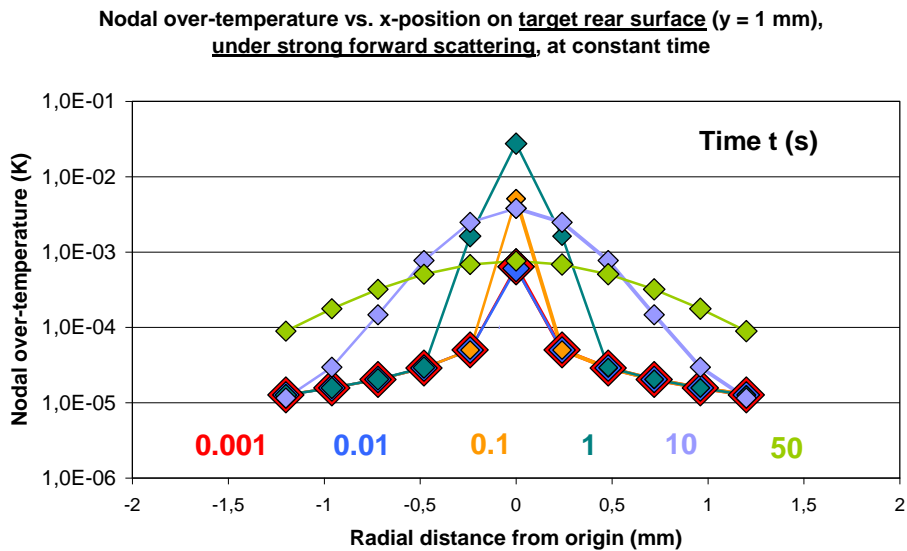

Figure 7b Nodal over-temperature of ZrO$_2$-sample on its *rear* surface, calculated as before (Figure 6b) but now under *strong forward scattering* from the Two Flux-Model integrated in the FE procedure, and for *small* optical thickness, τ = 1. Results are given at different radial distances, Δx (mm), from the origin (x = 0, y = 1 mm) within the target and pellet area, at given times, t.

## 5 Extraction of thermal diffusivity

### 5.1 Description of the method

Considering only a single position, x, on front or rear sample surface according to Figures 2b at least would be questionable, and according to Figures 5a,b and 7a,b even be meaningless. Instead, in the present paper, the recently introduced Front Face Flash method (compare [10] and [15-18]) shall be applied to also samples of small optical thickness As announced, it considers temperature *distributions* for extraction of thermal diffusivity, not only a single position on rear (or front) surface. The method therefore is applicable in case there is energy exchange with environment (convection, radiation) that may introduce *local* variations of surface temperature. It is of practical value also when only part of the target or rear surface area might be visible.

Again, solely for improving clarity, the present analysis is confined to isotropic (constant, independent of temperature) materials properties (conductivity, specific heat, extinction coefficient, albedo), i. e. to idealistic conditions. But the procedures in principle are straightforward (though becoming increasingly laborious) when analysis is performed for anisotropic, temperature or wave length-dependent parameters, or for layered samples.

In Step (1) of the method, thermal homogeneity, $T_0(x,t)$, at any constant depth y, preferentially at y = 0 (see below, step 2)

$$T_0(t) = [T(t)/t]/[dT(t)/dt] \qquad (2)$$

has to be checked. It would be useless to apply the Front Face Flash method (Step 2) to regions where $T_0(t)$ oscillates. Step (1) identifies the region of Fourier numbers, Fo = D t/L$^2$ (using 0.16 ≤ Fo ≤ 0.25) to yield the time interval during which the heating regime is regular and where Step (2) may be applied:

In Step (2), the diffusivity, D, is extracted from absolute temperatures, $T_i^k$,

$$D = \frac{\dfrac{T_i^k - T_i^{k-1}}{\Delta t}}{\dfrac{T_{i+1}^k - T_{i-1}^k}{2r_i \Delta r} + \dfrac{T_{i+1}^k - 2T_i^k + T_{i-1}^k}{(\Delta r)^2}} \qquad (3)$$

The $T_i^k$, with indices i and k, denote temperature measured at time, $t^k$, at position number, i, with $r = r_i$ at front sample surface (y = 0); co-ordinate r is parallel to x-direction. Time and radial co-ordinate intervals are defined by $\Delta t = t_k - t_{k-1} = t_{k+1} - t_k$, and $\Delta r = r_{i+1} - r_i$. The $r_{i+1}$ and $r_i$ have to be taken outside the target. Temperatures applied in Eq. (3) are nodal values.

With the fine mesh and correspondingly improved spatial resolution, divisions by zero in Eq. (3) are avoided.

## 5.2   Results

The extracted thermal diffusivity, D, is given in Tables 1 and 2, for large, medium and small optical thickness and different albedo, respectively. Literature values of $ZrO_2$ and SiC, also given in Table 1, column 2, are strongly different, by more than one order of magnitude.

**Tab. 1** Radial diffusivity, D, obtained from application of the Front Face Flash Method; (A) Medium ($\tau = 10$) and large optical thickness ($\tau = 50$); all results are from [10]

| Pellet material | Diffusivity [m² s⁻¹], from literature data | Diffusivity [m² s⁻¹], from Front Face Flash Method |
|---|---|---|
| $ZrO_2$, solely conduction | 5.751 10⁻⁷ | 5.780 10⁻⁷ |
| $ZrO_2$, conduction/radiation, E = 10⁴ [m⁻¹], $\tau = 10$ | | 5.941 10⁻⁷ |
| $ZrO_2$, conduction/radiation, E = 5 10⁴ [m⁻¹], $\tau = 50$ | | 6.107 10⁻⁷ |
| SiC, solely conduction | 1.619 10⁻⁵ | 1.762 10⁻⁵ |
| SiC, conduction/radiation, E = 10⁴ [m⁻¹], $\tau = 10$ | | 1.923 10⁻⁵ |
| SiC, conduction/radiation, E = 5 10⁴ [m⁻¹], $\tau = 50$ | | 1.921 10⁻⁵ |

For solely solid conduction (blue symbols in Table 1), there is good (SiC), and in case of $ZrO_2$ almost perfect, agreement between "experiment" and application of the Front Face Flash method. Trivially, this result depends on accuracy of both Front Face Flash method (column 3) and literature values (column 2). Accuracy of the Front Face Flash method relies on appropriate choice of the positions $r_i$, constant $\Delta r$ and $\Delta t$ in Eq. (3), which in practical situations might not easily be realised (part of the rear sample surface perhaps could not be visible for a detector). Literature values of thermal conductivity are authentic if there was *solely* conduction heat transfer in the corresponding experiments, without *any* radiation contribution, i. e. with bulk samples.

**Tab. 2** Radial diffusivity, D, obtained from application of the Front Face Flash Method; (B) Small optical thickness (t = 1), for different albedo

| Pellet material | Diffusivity [m² s⁻¹] from Front Face Flash Method |
|---|---|
| ZrO$_2$, conduction/radiation, E = 10³ [m⁻¹], t = 1, Albedo = 0.01 (strong absorption) | 6.02 10⁻⁷ |
| ZrO$_2$, conduction/radiation, E = 10³ [m⁻¹], t = 1, Albedo = 0.99 (strong scattering) | 6.12 10⁻⁷ |

A general expectation is that extracted diffusivity, like those in Table 1, should be closer to the case of solely conduction if optical thickness is large; radiative transport then will not substantially contribute to total heat transfer. However, the extracted diffusivity reported in Table 1 (red symbols) is by almost 20% larger in case radiation *is* modelled, even if the optical thickness is assumed as large. The explanation of this apparent discrepancy is as follows:

Variation of the diffusivity, against the pure solid conduction value, results from opening a second heat transfer "channel", radiation parallel to conduction. If radiative transport is strong (optical thickness is small), this channel will be responsible for most of total heat transfer, and little contribution is left for solid conduction. If on the other hand optical thickness is large (radiation less important), the overwhelming part of total heat transfer is shifted to the parallel conductive flow channel. To which extent these expectations are realised depends on magnitude of the diffusivity:

In the poorly conducting solid ZrO$_2$, increase of E leads to an increase of the extracted diffusivity. In the strongly conducting solid SiC, on the other hand, variations of E hardly induce variations of extracted diffusivity since solid conduction in this material is so strong that the competing radiation channel, under (reasonable) variations of E or Ω, plays only a minor role. Diffusivity then remains constant, as is observed in Table 1.

We may speak of "coupling" between both heat transfer channels. Coupling in case of $ZrO_2$ apparently is strong, otherwise variations of optical thickness would not lead to a significant shift of heat transfer (indicated by the extracted, increasing diffusivity) from one channel to the other. In SiC, however, the extracted diffusivity of this material is almost the same for both optical thicknesses. Coupling in this material accordingly is weak. Both transport channels in SiC then behave as if the other is not present. This is a situation typical for the well-known "additive approximation" (radiation can simply be added, as a "correction", to the conductive channel). Variation of optical thickness then has little influence on the amount that conduction contributes to total heat flow. The extracted diffusivity accordingly remains almost unchanged.

Provided we may interpret this situation in terms of coupling: Which mechanism, in both $ZrO_2$ and SiC, initialises "coupling"? It is the temperature profile in the samples: Simultaneous solutions of the Equation of Radiative Transfer and of the Energy Conservation Equation are coupled by the temperature field in the materials, compare standard volumes on radiative transfer, e. g. Siegel, Howell: Thermal radiation heat transfer, 1972 Edition, Sect. 19.3.3. This shall be exemplified in $ZrO_2$:

Magnitude and distribution of volume power sources, $Q_V(i,j)$, and accordingly, the temperature distribution in the sample, depend on optical thickness. The $Q_V(i,j)$ in the $ZrO_2$-sample are shown for optical thickness $τ = 50$ and $τ = 1$ in Figure 8a,b, respectively. The $Q_V(i,j)$ are explained solely by absorption and scattering of bundles (the Monte Carlo approach, not subject to solid conduction); they depend also on the energy losses that the bundles have experienced up to then.

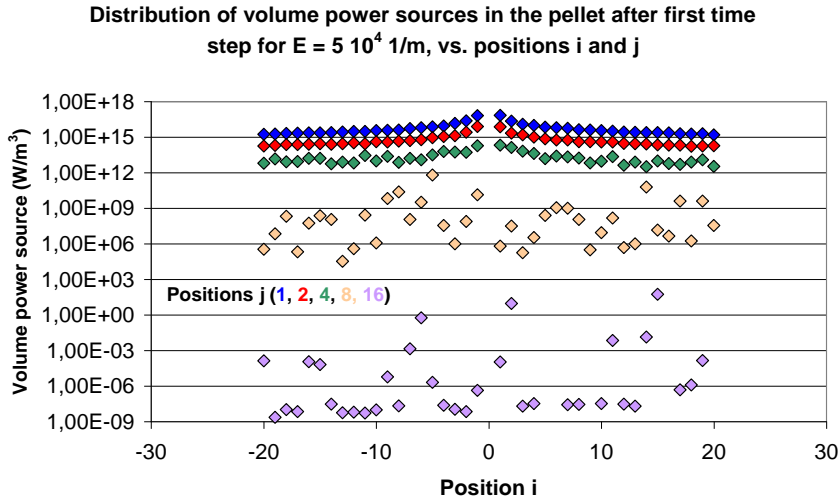

Figure 8a Magnitude and distribution of volume power sources, $Q_V(i,j)$ (W/m$^3$), obtained from the Monte Carlo-simulation of radiative transfer, for optical thickness τ = 50. Bundles are emitted randomly from target at positions located between -15 ≤ r ≤ 15 mm (their foot-points oscillate over the target area); the bundles subsequently are absorbed or scattered in the volume elements defined in Figure 3b. Data are given for different discrete co-ordinates i and j and at the end of the first time step (t = 1 ns) of the simulated (short) radiation pulsed (5 10$^4$ bundles) and for constant (independent of wavelength) extinction coefficient, E = 5 10$^4$ [m$^{-1}$]. Also albedo of single scattering, Ω, and anisotropy factor, $m_S$, are assumed as constant, $Ω_c$ = 0.5, $m_S$ = 2. The figure is taken from [10].

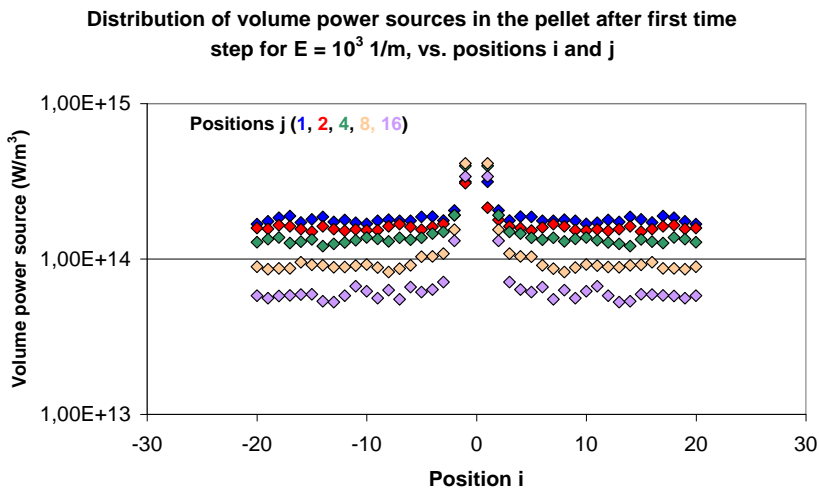

Figure 8b Magnitude and distribution of volume power sources, $Q_V(i,j)$ (W/m$^3$), obtained from the Monte Carlo-simulation of radiative transfer. Results are calculated as before (Figure 8a), but for small optical thickness, τ = 1. The figure is taken from [10].

If optical thickness is large (Figure 8a), the $Q_V(i,j)$ near front surface of the $ZrO_2$-pellet (j ≤ 4), and at positions close to the axis of symmetry (x = 0), are very large, and their distribution (variation with co-ordinate y) becomes more and more diffuse. This means the radiation field resulting from emission by all volume elements (i,j) will become more and more isotropic.

If on the other hand the optical thickness is small (Figure 8b), the distribution of the $Q_V(i,j)$ is more uniform, and large values will no longer be observed solely close to the sample front side, but also deeply below. Second, the number of bundles escaping from the sample increases. Both observations reduce overall volume power density, in Figure 8b by roughly two orders of magnitude against Figure 8a. As a consequence, the temperature profiles within the $ZrO_2$-sample (Figure 9a,b) are expected to look very different:

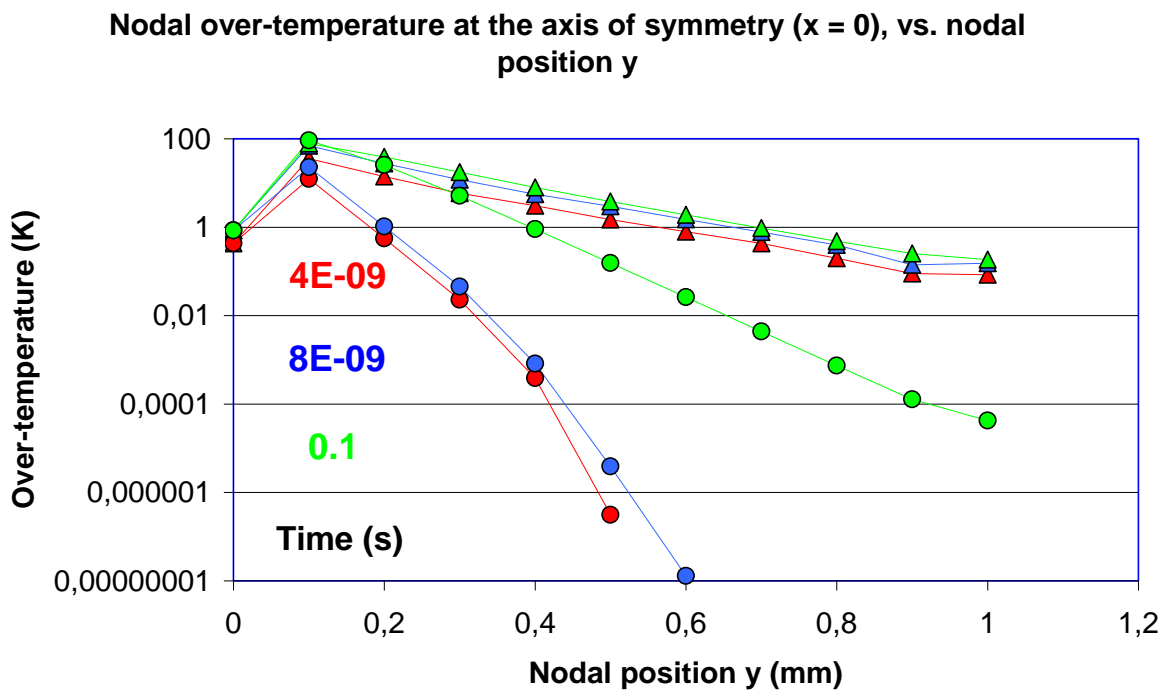

Figure 9a Temperature distribution (nodal over-temperatures) within the $ZrO_2$-sample along the axis of symmetry (x = 0, 0 ≤ y ≤ L) for large optical thickness, τ = 10 (triangles) and τ = 50 (circles) at given time, t, vs. nodal position, y. Data are from [10].

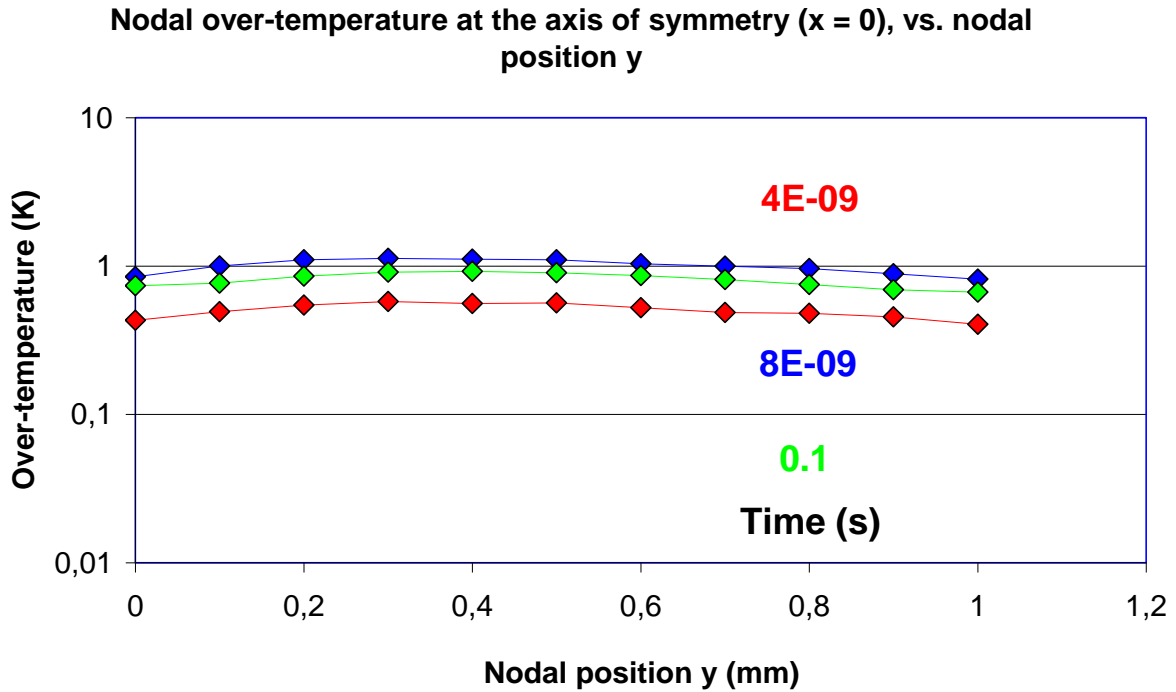

Figure 9b Temperature distribution (nodal over-temperatures) of the $ZrO_2$-sample along the axis of symmetry ($x = 0$, $0 \leq y \leq L$) for small optical thickness $\tau = 1$ at given time, t, vs. nodal position, y. Data are from [10].

An approximately linear decay of T(y,t) (note the logarithmic scale) with increasing co-ordinate y, is observed in Figure 9a for $\tau = 50$ and $\tau = 10$, with a strong local temperature maximum near the sample front surface; it is due to the volume heat sources $Q_V(i,j)$. Large temperature near y = 0 (front surface) causes radiation to dominate at these positions, while reduced temperature observed at large y (near rear surface) indicates reduced radiation contribution. By conservation of energy, solid conduction increases at these positions; compare e. g. the frequently referenced work of Viskanta [9]. Coupling between both channels thus is strong.

For small optical thickness, the temperature distribution is strongly different: The field T(y,t) in Figure 9b, along the axis of symmetry (x = 0), becomes more and more uniform when optical thickness in the simulations is reduced to finally $\tau = 1$. The almost constant T(y,t) indicates that the interrelation between radiation and conduction heat transfer, under small optical thickness, is weak, and radiation dominates so strongly that substantial variations of (extracted) solid diffusivity cannot be seen.

To which extent, in experimental situations, a shift of proper solid to an *apparent* (conductive/radiative) diffusivity can be realised accordingly depends on materials radiative properties (extinction coefficient E, albedo Ω) and on sample thickness, L. The question is how the extracted diffusivity then has to be interpreted. Is it still a materials property?

## 6    Conclusions and potential applicability of the method to thin film superconductors

Assume again that the pellet material has isotropic thermal materials properties. If we consider only 1D conduction heat flow, along the y-direction in Figure 3b, the energy balance, in its most simple form (without extra heat sources and sinks), reads

$$\rho\, c_p\, \partial T/\partial t = \mathrm{div}(\mathbf{q}) = \mathbf{div}\{\lambda\, \mathbf{grad}\,[T(y)]\} \tag{4a}$$

with ρ, $c_p$ and λ the density, specific heat and thermal conductivity, respectively, and with **q** the heat flux, **q** = λ **grad**(T); bold letters indicate vectors. Eq. (4a) reduces to

$$\rho\, c_p\, \partial T/\partial t = \lambda\, [\mathbf{div}\,(\mathbf{grad}(T)] \tag{4b}$$

provided the thermal conductivity is constant (does not depend on temperature field, T(y,t) or on materials in-homogeneities). The temperature field accordingly must be differentiable with respect to co-ordinate y, otherwise its gradient would not exist. In turn, if a missing gradient would be substituted by ratios ΔT/Δy of finite temperature and co-ordinate differences, thermal conductivity, in order to fulfil the energy balance (left side of Eq. 4a,b kept constant),

(i) either depends on how exact the ratio ΔT/Δy approximates the actual temperature variation in the pellet. Thermal conductivity, λ, then is no longer a solely materials property but an apparent conductivity, *or*

(ii) if the thermal conductivity is kept constant, the left side of Eq. (4a,b), i. e. ∂T/∂t, the temperature excursion with time, would depend on the approximation ΔT/Δy.

Situations (i) and (ii) are different from the procedure reported in the conduction/radiation analysis by Viskanta [9]: There, a thermal conductivity was explicitly pre-specified. The question then is under which circumstances a temperature field, T(y), would not be differentiable.

This occurs if optical thickness of the sample is small, τ ≤ 1. The mean free path, $l_m$, in direction parallel to the surface normal, then is statistically larger than sample thickness, which means a large number of bundles within the pellet might not find absorbing or scattering centres for radiation/solid particle collisions. The centres, at least part of their total number, even might not have solid/solid contacts among themselves, and a temperature field between these centres accordingly is not defined. Local or non-local temperature gradients then cannot exist. Thermal conductivity, in samples of small optical thickness, thus is at best an apparent conductivity but no longer a uniquely defined materials property. The same applies to the thermal diffusivity, $D = \lambda/(\rho\, c_p)$.

The Front Face Flash-model for extracting thermal diffusivity from surface temperature distributions was derived by the late Oleg Yu. Troitsky, the co-author of this paper, before 1999. The method entirely relies on the imagination of *solely thermal conduction*, including no radiation at all, which means application of the original model to thin films with small optical thickness would deliver anything else than an apparent conductivity. Strictly speaking, even the name "Diffusivity" would be misleading because it is no longer a diffusive energy transport but rather an *exchange* of radiation between solid particles (provided they exist in thin film sampls), parallel to solid conduction. Thermal transport in such cases can tentatively be modelled with cell models but it is questionable whether a *coupled* conduction/radiation problem can be solved accurately.

But optical thickness τ << 1 is typical for superconductor, thin film samples: Thickness of thin ceramic, high temperature superconductor films is 1 or 2 micrometers, with E in the order $10^4$ to $10^5$ 1/m. Investigating the applicability of the Front Face Flash method to the corresponding multi-layer "coated conductor" system (Figure 10, with substrate, buffer layer, thin film superconductor and metallic jacket for conductor stability) thus is most challenging.

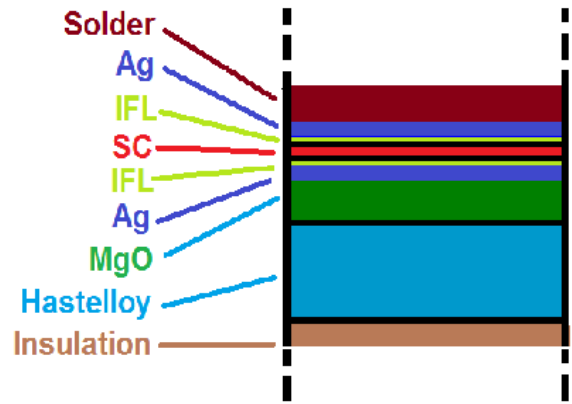

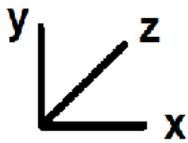

Figure 10 Cross section (schematic) showing conductor geometry and composition of a coated, multilayer YBaCuO3 123 thin film superconductor. Crystallographic c-axis of the YBaCuO-layer is parallel to y-axis of the co-ordinate system. Conductor architecture and dimensions are standard. Superconductor (SC) layer thickness (red sections) is 2 µm, width 6 mm; thickness and width of the Ag elements (lilac) is the same, width of the interfacial layers (IFL, light green) is approximately 40 nm (the IFL are included to simulate surface roughness and diffusion of species between the SC and its neighbouring Ag- and MgO-layers, respectively).

Knowledge of the diffusivity is highly important for design of superconductors to be stable against quench. The Front Face Flash Method would be suitable to determine the diffusivity of the total conductor thickness including all its components. This would be sufficient for solely the thermal design aspect. Determination of the diffusivity of solely the single SC layer probably would have to apply other methods (non-stationary, like 3ω).

**Appendix**

**A) Monte Carlo simulations**

The scattering phase function is a random variable and thus is different at each of interactions between bundles and solid constituents (the constituents are simulated by the volume elements); this starts with the target elements.

We use the following spider diagram:

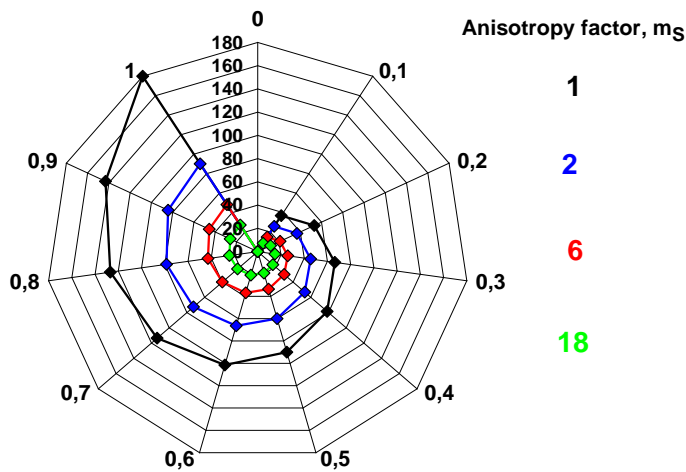

The diagram shows the angular distribution of bundles remitted from any volume element (including those of the target) in dependence of an anisotropy factor, $m_S$ (see below for its definition), and of the random variable, $0 \leq R_\beta \leq 1$ (volume elements, as mentioned, are created from the area elements in Figure 3b by rotation against the x = 0 axis). Emission from especially the target elements is simulated with random foot-point co-ordinates of the bundles; this is a much better (and a more realistic) approximation than naively assuming that a single beam would leave the target just at the co-ordinate origin (x = 0, y = 0). In other words, the foot-points oscillate over the target area.

The diagram illustrates that remission or scattering angles, $0 \leq \beta \leq 180°$, are the smaller the larger $m_S$. The innermost black circle denotes $\beta = 0$. For example, if $m_S = 16$ and $R_\beta = 1$, the cone angle amounts to $\beta < 27.3°$ against normal to the surface. Remission or scattering at still smaller cone angles is less probable, $R_\beta < 1$ (follow the red circle counter clock-wise).

In case of isotropic scattering, the angular distribution of bundles has a cone angle $\beta$ that scales with

$$\sin(\beta) = (R_\beta)^{1/2} \tag{A1}$$

Anisotropic scattering then is modeled by division of the random number, $R_\beta$, by a constant, $m_S$. Large $m_S$ introduced in

$$\beta = \arccos(1 - 2R_\beta / m_S) \tag{A2}$$

provides strongly forward peaked angular distributions. This is applied to each scattering process in each of the volume elements.

For simplicity, the Monte Carlo simulations were performed with constant (independent of wavelength) radiative properties, like extinction coefficients, $E$ ($10^3$, $10^4$ and $5 \cdot 10^4$ [$m^{-1}$]), albedo $\Omega = 0.5$ (or $\Omega = 0.01$ and $0.99$, to simulate extreme cases), anisotropy coefficient $m_S = 2$, index of refraction $n = 2$, for both $ZrO_2$ and SiC (and, for comparison, $\Omega_c = 0.02$, $m_S = 6$, $n = 2$ for graphite in [10], and also with wave length-dependent parameters).

These optical constants are approximations only but rely on quantitative estimates from Mie theory of scattering, from corresponding approximations of extinction cross sections reported by van de Hulst and on experimental work of Cabannes and Billard. The procedures are described in detail in [10] with references to the original work, and in our previous papers on thin ceramic and superconductor films.

### B) Finite Element calculations

We have used the same isotropic thermal materials properties as in previous papers:

| Pellet material, reference | ZrO$_2$ (a) | SiC (b) |
|---|---|---|
| Thermal conductivity, λ [W m$^{-1}$ K$^{-1}$] | 1.675 isotropic | 43.5 isotropic |
| Density [kg m$^{-3}$] | 5700 | 3210 |
| Specific heat [Ws kg$^{-1}$ K$^{-1}$] | 511 | 837 |

with (a) taken from [19], p. 267, (b) [19], p. 925-927.